# The Author-Topic Model for Authors and Documents

**Michal Rosen-Zvi**
Dept. of Computer Science
UC Irvine
Irvine, CA 92697-3425, USA

**Thomas Griffiths**
Dept. of Psychology
Stanford University
Stanford, CA 94305, USA

**Mark Steyvers**
Dept. of Cognitive Sciences
UC Irvine
Irvine, CA 92697, USA

**Padhraic Smyth**
Dept. of Computer Science
UC Irvine
Irvine, CA 92697-3425, USA

## Abstract

We introduce the author-topic model, a generative model for documents that extends Latent Dirichlet Allocation (LDA; Blei, Ng, & Jordan, 2003) to include authorship information. Each author is associated with a multinomial distribution over topics and each topic is associated with a multinomial distribution over words. A document with multiple authors is modeled as a distribution over topics that is a mixture of the distributions associated with the authors. We apply the model to a collection of 1,700 NIPS conference papers and 160,000 CiteSeer abstracts. Exact inference is intractable for these datasets and we use Gibbs sampling to estimate the topic and author distributions. We compare the performance with two other generative models for documents, which are special cases of the author-topic model: LDA (a topic model) and a simple author model in which each author is associated with a distribution over words rather than a distribution over topics. We show topics recovered by the author-topic model, and demonstrate applications to computing similarity between authors and entropy of author output.

## 1 Introduction

Characterizing the content of documents is a standard problem addressed in information retrieval, statistical natural language processing, and machine learning. A representation of document content can be used to organize, classify, or search a collection of documents. Recently, generative models for documents have begun to explore topic-based content representations, modeling each document as a mixture of probabilistic topics (e.g., Blei, Ng, & Jordan, 2003; Hofmann, 1999).

Here, we consider how these approaches can be used to address another fundamental problem raised by large document collections: modeling the interests of authors.

By modeling the interests of authors, we can answer a range of important queries about the content of document collections. With an appropriate author model, we can establish which subjects an author writes about, which authors are likely to have written documents similar to an observed document, and which authors produce similar work. However, research on author modeling has tended to focus on the problem of authorship attribution – who wrote which document – for which discriminative models based on relatively superficial features are often sufficient. For example, the "stylometric" approach (e.g., Holmes & Forsyth, 1995) finds stylistic features (e.g., frequency of certain stop words, sentence lengths, diversity of an author's vocabulary) that discriminate between different authors.

In this paper we describe a generative model for document collections, the author-topic model, that simultaneously models the content of documents and the interests of authors. This generative model represents each document with a mixture of topics, as in state-of-the-art approaches like Latent Dirichlet Allocation (Blei et al., 2003), and extends these approaches to author modeling by allowing the mixture weights for different topics to be determined by the authors of the document. By learning the parameters of the model, we obtain the set of topics that appear in a corpus and their relevance to different documents, as well as identifying which topics are used by which authors.

The paper is organized as follows. In Section 2, we discuss generative models for documents using authors and topics, and introduce the author-topic model. We devote Section 3 to describing the Gibbs sampler used for inferring the model parameters, and in Section 4 we present the results of applying this algorithm to two collections of computer science documents—NIPS conference papers and abstracts from the CiteSeer database. We conclude and discuss further research directions in Section 5.

## 2 Generative models for documents

We will describe three generative models for documents: one that models documents as a mixture of topics (Blei et al., 2003), one that models authors with distributions over words, and one that models both authors and documents using topics. All three models use the same notation. A document $d$ is a vector of $N_d$ words, $\mathbf{w}_d$, where each $w_{id}$ is chosen from a vocabulary of size $V$, and a vector of $A_d$ authors $\mathbf{a}_d$, chosen from a set of authors of size $A$. A collection of $D$ documents is defined by $\mathcal{D} = \{(\mathbf{w}_1, \mathbf{a}_1), \ldots, (\mathbf{w}_D, \mathbf{a}_D)\}$.

### 2.1 Modeling documents with topics

A number of recent approaches to modeling document content are based upon the idea that the probability distribution over words in a document can be expressed as a mixture of topics, where each topic is a probability distribution over words (e.g., Blei, et al., 2003; Hofmann, 1999). We will describe one such model – Latent Dirichlet Allocation (LDA; Blei et al., 2003).[1] In LDA, the generation of a document collection is modeled as a three step process. First, for each document, a distribution over topics is sampled from a Dirichlet distribution. Second, for each word in the document, a single topic is chosen according to this distribution. Finally, each word is sampled from a multinomial distribution over words specific to the sampled topic.

This generative process corresponds to the hierarchical Bayesian model shown (using plate notation) in Figure 1(a). In this model, $\phi$ denotes the matrix of topic distributions, with a multinomial distribution over $V$ vocabulary items for each of $T$ topics being drawn independently from a symmetric Dirichlet($\beta$) prior. $\theta$ is the matrix of document-specific mixture weights for these $T$ topics, each being drawn independently from a symmetric Dirichlet($\alpha$) prior. For each word, $z$ denotes the topic responsible for generating that word, drawn from the $\theta$ distribution for that document, and $w$ is the word itself, drawn from the topic distribution $\phi$ corresponding to $z$. Estimating $\phi$ and $\theta$ provides information about the topics that participate in a corpus and the weights of those topics in each document respectively. A variety of algorithms have been used to estimate these parameters, including variational inference (Blei et al., 2003), expectation propagation (Minka & Lafferty, 2002), and Gibbs sampling (Griffiths & Steyvers, 2004). However, this topic model provides no explicit information about the interests of authors: while it is informative about the content of documents, authors may produce several documents – often with co-authors – and it is consequently unclear how the topics used in these documents might be used to describe the interests of the authors.

### 2.2 Modeling authors with words

Topic models illustrate how documents can be modeled as mixtures of probability distributions. This suggests a simple method for modeling the interests of authors. Assume that a group of authors, $\mathbf{a}_d$, decide to write the document $d$. For each word in the document an author is chosen uniformly at random, and a word is chosen from a probability distribution over words that is specific to that author.

This model is similar to a mixture model proposed by McCallum (1999) and is equivalent to a variant of LDA in which the mixture weights for the different topics are fixed. The underlying graphical model is shown in Figure 1(b). $x$ indicates the author of a given word, chosen uniformly from the set of authors $\mathbf{a}_d$. Each author is associated with a probability distribution over words $\phi$, generated from a symmetric Dirichlet($\beta$) prior. Estimating $\phi$ provides information about the interests of authors, and can be used to answer queries about author similarity and authors who write on subjects similar to an observed document. However, this author model does not provide any information about document content that goes beyond the words that appear in the document and the authors of the document.

### 2.3 The author-topic model

The author-topic model draws upon the strengths of the two models defined above, using a topic-based representation to model both the content of documents and the interests of authors. As in the author model, a group of authors, $\mathbf{a}_d$, decide to write the document $d$. For each word in the document an author is chosen uniformly at random. Then, as in the topic model, a topic is chosen from a distribution over topics specific to that author, and the word is generated from the chosen topic.

The graphical model corresponding to this process is shown in Figure 1(c). As in the author model, $x$ indicates the author responsible for a given word, chosen from $\mathbf{a}_d$. Each author is associated with a distribution over topics, $\theta$, chosen from a symmetric Dirichlet($\alpha$) prior. The mixture weights corresponding to the chosen author are used to select a topic $z$, and a word is generated according to the distribution $\phi$ corresponding to that topic, drawn from a symmetric Dirichlet($\beta$) prior.

---

[1] The model we describe is actually the *smoothed* LDA model (Blei et al., 2003) with symmetric Dirichlet priors (Griffiths & Steyvers, 2004) as this is closest to the author-topic model.

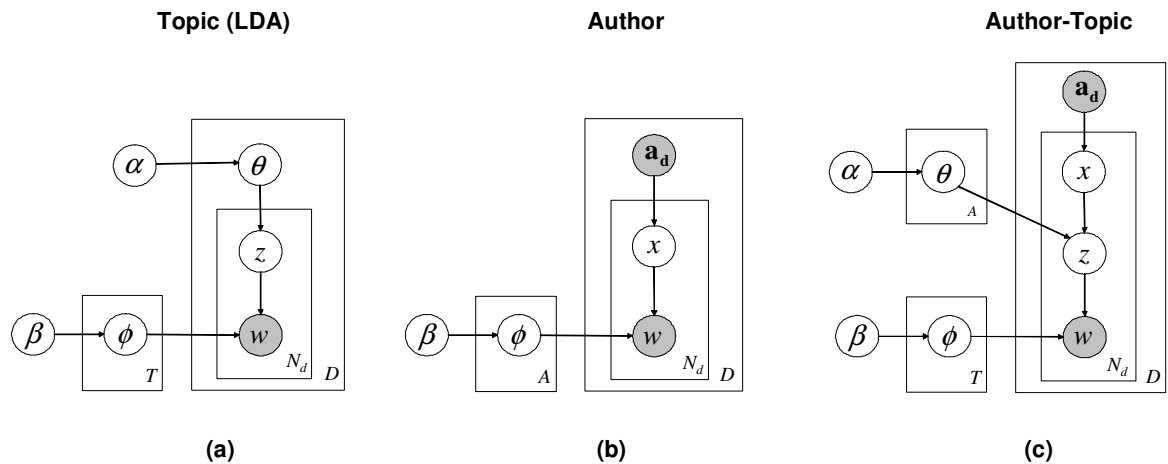

Figure 1: Generative models for documents. (a) Latent Dirichlet Allocation (LDA; Blei et al., 2003), a topic model. (b) An author model. (c) The author-topic model.

The author-topic model subsumes the two models described above as special cases: topic models like LDA correspond to the case where each document has one unique author, and the author model corresponds to the case where each author has one unique topic. By estimating the parameters $\phi$ and $\theta$, we obtain information about which topics authors typically write about, as well as a representation of the content of each document in terms of these topics. In the remainder of the paper, we will describe a simple algorithm for estimating these parameters, compare these different models, and illustrate how the results produced by the author-topic model can be used to answer questions about which which authors work on similar topics.

## 3 Gibbs sampling algorithms

A variety of algorithms have been used to estimate the parameters of topic models, from basic expectation-maximization (EM; Hofmann, 1999), to approximate inference methods like variational EM (Blei et al., 2003), expectation propagation (Minka & Lafferty, 2002), and Gibbs sampling (Griffiths & Steyvers, 2004). Generic EM algorithms tend to face problems with local maxima in these models (Blei et al., 2003), suggesting a move to approximate methods in which some of the parameters—such as $\phi$ and $\theta$—can be integrated out rather than explicitly estimated. In this paper, we will use Gibbs sampling, as it provides a simple method for obtaining parameter estimates under Dirichlet priors and allows combination of estimates from several local maxima of the posterior distribution.

The LDA model has two sets of unknown parameters – the $D$ document distributions $\theta$, and the $T$ topic distributions $\phi$ – as well as the latent variables corresponding to the assignments of individual words to topics $z$. By applying Gibbs sampling (see Gilks, Richardson, & Spiegelhalter, 1996), we construct a Markov chain that converges to the posterior distribution on $z$ and then use the results to infer $\theta$ and $\phi$ (Griffiths & Steyvers, 2004). The transition between successive states of the Markov chain results from repeatedly drawing $z$ from its distribution conditioned on all other variables, summing out $\theta$ and $\phi$ using standard Dirichlet integrals:

$$P(z_i = j | w_i = m, \mathbf{z}_{-i}, \mathbf{w}_{-i}) \propto \frac{C_{mj}^{WT} + \beta}{\sum_{m'} C_{m'j}^{WT} + V\beta} \frac{C_{dj}^{DT} + \alpha}{\sum_{j'} C_{dj'}^{DT} + T\alpha} \quad (1)$$

where $z_i = j$ represents the assignments of the $i$th word in a document to topic $j$, $w_i = m$ represents the observation that the $i$th word is the $m$th word in the lexicon, and $\mathbf{z}_{-i}$ represents all topic assignments not including the $i$th word. Furthermore, $C_{mj}^{WT}$ is the number of times word $m$ is assigned to topic $j$, not including the current instance, and $C_{dj}^{DT}$ is the number of times topic $j$ has occurred in document $d$, not including the current instance. For any sample from this Markov chain, being an assignment of every word to a topic, we can estimate $\phi$ and $\theta$ using

$$\phi_{mj} = \frac{C_{mj}^{WT} + \beta}{\sum_{m'} C_{m'j}^{WT} + V\beta} \quad (2)$$

$$\theta_{dj} = \frac{C_{dj}^{DT} + \alpha}{\sum_{j'} C_{dj'}^{DT} + T\alpha} \quad (3)$$

where $\phi_{mj}$ is the probability of using word $m$ in topic $j$, and $\theta_{dj}$ is the probability of topic $j$ in document $d$. These values correspond to the predictive distributions over new words $w$ and new topics $z$ conditioned on $\mathbf{w}$ and $\mathbf{z}$.

An analogous approach can be used to derive a Gibbs sampler for the author model. Specifically, we have

$$P(x_i = k | w_i = m, \mathbf{x}_{-i}, \mathbf{w}_{-i}, \mathbf{a}_d) \propto \frac{C_{mk}^{WA} + \beta}{\sum_{m'} C_{m'k}^{WA} + V\beta}$$

where $x_i = k$ represents the assignments of the $i$th word in a document to author $k$ and $C_{mk}^{WA}$ is the number of times word $m$ is assigned to author $k$. An estimate of $\phi$ can be obtained via

$$\phi_{mk} = \frac{C_{mk}^{WA} + \beta}{\sum_{m'} C_{m'k}^{WA} + V\beta}$$

similar to Equation 2.

In the author-topic model, we have two sets of latent variables: $z$ and $x$. We draw each $(z_i, x_i)$ pair as a block, conditioned on all other variables:

$$P(z_i = j, x_i = k | w_i = m, \mathbf{z}_{-i}, \mathbf{x}_{-i}, \mathbf{w}_{-i}, \mathbf{a}_d) \propto$$
$$\frac{C_{mj}^{WT} + \beta}{\sum_{m'} C_{m'j}^{WT} + V\beta} \frac{C_{kj}^{AT} + \alpha}{\sum_{j'} C_{kj'}^{AT} + T\alpha} \quad (4)$$

where $z_i = j$ and $x_i = k$ represent the assignments of the $i$th word in a document to topic $j$ and author $k$ respectively, $w_i = m$ represents the observation that the $i$th word is the $m$th word in the lexicon, and $\mathbf{z}_{-i}, \mathbf{x}_{-i}$ represent all topic and author assignments not including the $i$th word, and $C_{kj}^{AT}$ is the number of times author $k$ is assigned to topic $j$, not including the current instance. Equation 4 is the conditional probability derived by marginalizing out the random variables $\phi$ (the probability of a word given a topic) and $\theta$ (the probability of a topic given an author). These random variables are estimated from samples via

$$\phi_{mj} = \frac{C_{mj}^{WT} + \beta}{\sum_{m'} C_{m'j}^{WT} + V\beta} \quad (5)$$

$$\theta_{kj} = \frac{C_{kj}^{AT} + \alpha}{\sum_{j'} C_{kj'}^{AT} + T\alpha} \quad (6)$$

In the examples considered here, we do not estimate the hyperparameters $\alpha$ and $\beta$—instead the smoothing parameters are fixed at $50/T$ and 0.01 respectively.

Each of these algorithms requires tracking only small amounts of information from a corpus. For example, in the author-topic model, the algorithm only needs to keep track of a $V \times T$ (word by topic) count matrix, and an $A \times T$ (author by topic) count matrix, both of which can be represented efficiently in sparse format. We start the algorithm by assigning words to random topics and authors (from the set of authors on the document). Each iteration of the algorithm involves applying Equation 4 to every word token in the document collection, which leads to a time complexity that is of order of the total number of word tokens in the training data set multiplied by the number of topics, $T$ (assuming that the number of authors on each document has negligible contribution to the complexity). The count matrices are saved at the 2000th iteration of this sampling process. We do this 10 times so that 10 samples are collected in this manner (the Markov chain is started 10 times from random initial assignments).

## 4  Experimental results

In our results we used two text data sets consisting of technical papers—full papers from the NIPS conference[2] and abstracts from CiteSeer (Lawrence, Giles, & Bollacker, 1999). We removed extremely common words from each corpus, a standard procedure in "bag of words" models. This leads to a vocabulary size of $V = 13,649$ unique words in the NIPS data set and $V = 30,799$ unique words in the CiteSeer data set. Our collection of NIPS papers contains $D = 1,740$ papers with $K = 2,037$ authors and a total of $2,301,375$ word tokens. Our collection of CiteSeer abstracts contains $D = 162,489$ abstracts with $K = 85,465$ authors and a total of $11,685,514$ word tokens.

### 4.1  Examples of topic and author distributions

The NIPS data set contains papers from the NIPS conferences between 1987 and 1999. The conference is characterized by contributions from a number of different research communities in the general area of learning algorithms. Figure 2 illustrates examples of 8 topics (out of 100) as learned by the model for the NIPS corpus. The topics are extracted from a single sample at the 2000th iteration of the Gibbs sampler. Each topic is illustrated with (a) the top 10 words most likely to be generated conditioned on the topic, and (b) the top 10 most likely authors to have generated a word conditioned on the topic. The first 6 topics we selected for display (left to right across the top and the first two on the left on the bottom) are quite specific representations of different topics that have been popular at the NIPS conference over the time-period 1987–99: EM and mixture models, handwritten character recognition, reinforcement learning, SVMs and kernel methods, speech recognition, and Bayesian learning.

---

[2]The NIPS data set in Matlab format is available online at http://www.cs.toronto.edu/~roweis/data.html.

| TOPIC 19 | | TOPIC 24 | | TOPIC 29 | | TOPIC 87 | |
|---|---|---|---|---|---|---|---|
| WORD | PROB. | WORD | PROB. | WORD | PROB. | WORD | PROB. |
| LIKELIHOOD | 0.0539 | RECOGNITION | 0.0400 | REINFORCEMENT | 0.0411 | KERNEL | 0.0683 |
| MIXTURE | 0.0509 | CHARACTER | 0.0336 | POLICY | 0.0371 | SUPPORT | 0.0377 |
| EM | 0.0470 | CHARACTERS | 0.0250 | ACTION | 0.0332 | VECTOR | 0.0257 |
| DENSITY | 0.0398 | TANGENT | 0.0241 | OPTIMAL | 0.0208 | KERNELS | 0.0217 |
| GAUSSIAN | 0.0349 | HANDWRITTEN | 0.0169 | ACTIONS | 0.0208 | SET | 0.0205 |
| ESTIMATION | 0.0314 | DIGITS | 0.0159 | FUNCTION | 0.0178 | SVM | 0.0204 |
| LOG | 0.0263 | IMAGE | 0.0157 | REWARD | 0.0165 | SPACE | 0.0188 |
| MAXIMUM | 0.0254 | DISTANCE | 0.0153 | SUTTON | 0.0164 | MACHINES | 0.0168 |
| PARAMETERS | 0.0209 | DIGIT | 0.0149 | AGENT | 0.0136 | REGRESSION | 0.0155 |
| ESTIMATE | 0.0204 | HAND | 0.0126 | DECISION | 0.0118 | MARGIN | 0.0151 |
| AUTHOR | PROB. | AUTHOR | PROB. | AUTHOR | PROB. | AUTHOR | PROB. |
| Tresp_V | 0.0333 | Simard_P | 0.0694 | Singh_S | 0.1412 | Smola_A | 0.1033 |
| Singer_Y | 0.0281 | Martin_G | 0.0394 | Barto_A | 0.0471 | Scholkopf_B | 0.0730 |
| Jebara_T | 0.0207 | LeCun_Y | 0.0359 | Sutton_R | 0.0430 | Burges_C | 0.0489 |
| Ghahramani_Z | 0.0196 | Denker_J | 0.0278 | Dayan_P | 0.0324 | Vapnik_V | 0.0431 |
| Ueda_N | 0.0170 | Henderson_D | 0.0256 | Parr_R | 0.0314 | Chapelle_O | 0.0210 |
| Jordan_M | 0.0150 | Revow_M | 0.0229 | Dietterich_T | 0.0231 | Cristianini_N | 0.0185 |
| Roweis_S | 0.0123 | Platt_J | 0.0226 | Tsitsiklis_J | 0.0194 | Ratsch_G | 0.0172 |
| Schuster_M | 0.0104 | Keeler_J | 0.0192 | Randlov_J | 0.0167 | Laskov_P | 0.0169 |
| Xu_L | 0.0098 | Rashid_M | 0.0182 | Bradtke_S | 0.0161 | Tipping_M | 0.0153 |
| Saul_L | 0.0094 | Sackinger_E | 0.0132 | Schwartz_A | 0.0142 | Sollich_P | 0.0141 |

| TOPIC 31 | | TOPIC 61 | | TOPIC 71 | | TOPIC 100 | |
|---|---|---|---|---|---|---|---|
| WORD | PROB. | WORD | PROB. | WORD | PROB. | WORD | PROB. |
| SPEECH | 0.0823 | BAYESIAN | 0.0450 | MODEL | 0.4963 | HINTON | 0.0329 |
| RECOGNITION | 0.0497 | GAUSSIAN | 0.0364 | MODELS | 0.1445 | VISIBLE | 0.0124 |
| HMM | 0.0234 | POSTERIOR | 0.0355 | MODELING | 0.0218 | PROCEDURE | 0.0120 |
| SPEAKER | 0.0226 | PRIOR | 0.0345 | PARAMETERS | 0.0205 | DAYAN | 0.0114 |
| CONTEXT | 0.0224 | DISTRIBUTION | 0.0259 | BASED | 0.0116 | UNIVERSITY | 0.0114 |
| WORD | 0.0166 | PARAMETERS | 0.0199 | PROPOSED | 0.0103 | SINGLE | 0.0111 |
| SYSTEM | 0.0151 | EVIDENCE | 0.0127 | OBSERVED | 0.0100 | GENERATIVE | 0.0109 |
| ACOUSTIC | 0.0134 | SAMPLING | 0.0117 | SIMILAR | 0.0083 | COST | 0.0106 |
| PHONEME | 0.0131 | COVARIANCE | 0.0117 | ACCOUNT | 0.0069 | WEIGHTS | 0.0105 |
| CONTINUOUS | 0.0129 | LOG | 0.0112 | PARAMETER | 0.0068 | PARAMETERS | 0.0096 |
| AUTHOR | PROB. | AUTHOR | PROB. | AUTHOR | PROB. | AUTHOR | PROB. |
| Waibel_A | 0.0936 | Bishop_C | 0.0563 | Omohundro_S | 0.0088 | Hinton_G | 0.2202 |
| Makhoul_J | 0.0238 | Williams_C | 0.0497 | Zemel_R | 0.0084 | Zemel_R | 0.0545 |
| De-Mori_R | 0.0225 | Barber_D | 0.0368 | Ghahramani_Z | 0.0076 | Dayan_P | 0.0340 |
| Bourlard_H | 0.0216 | MacKay_D | 0.0323 | Jordan_M | 0.0075 | Becker_S | 0.0266 |
| Cole_R | 0.0200 | Tipping_M | 0.0216 | Sejnowski_T | 0.0071 | Jordan_M | 0.0190 |
| Rigoll_G | 0.0191 | Rasmussen_C | 0.0215 | Atkeson_C | 0.0070 | Mozer_M | 0.0150 |
| Hochberg_M | 0.0176 | Opper_M | 0.0204 | Bower_J | 0.0066 | Williams_C | 0.0099 |
| Franco_H | 0.0163 | Attias_H | 0.0155 | Bengio_Y | 0.0062 | de-Sa_V | 0.0087 |
| Abrash_V | 0.0157 | Sollich_P | 0.0143 | Revow_M | 0.0059 | Schraudolph_N | 0.0078 |
| Movellan_J | 0.0149 | Schottky_B | 0.0128 | Williams_C | 0.0054 | Schmidhuber_J | 0.0056 |

Figure 2: An illustration of 8 topics from a 100-topic solution for the NIPS collection. Each topic is shown with the 10 words and authors that have the highest probability conditioned on that topic.

For each topic, the top 10 most likely authors are well-known authors in terms of NIPS papers written on these topics (e.g., Singh, Barto, and Sutton in reinforcement learning). While most (order of 80 to 90%) of the 100 topics in the model are similarly specific in terms of semantic content, the remaining 2 topics we display illustrate some of the other types of "topics" discovered by the model. Topic 71 is somewhat generic, covering a broad set of terms typical to NIPS papers, with a somewhat flatter distribution over authors compared to other topics. Topic 100 is somewhat oriented towards Geoff Hinton's group at the University of Toronto, containing the words that commonly appeared in NIPS papers authored by members of that research group, with an author list largely consisting of Hinton plus his past students and postdocs.

Figure 3 shows similar types of results for 4 selected topics from the CiteSeer data set, where again topics on speech recognition and Bayesian learning show up. However, since CiteSeer is much broader in content (covering computer science in general) compared to NIPS, it also includes a large number of topics not seen in NIPS, from user interfaces to solar astrophysics (Figure 3). Again the author lists are quite sensible—for example, Ben Shneiderman is a widely-known senior figure in the area of user-interfaces.

| TOPIC 10 | | TOPIC 209 | | TOPIC 87 | | TOPIC 20 | |
|---|---|---|---|---|---|---|---|
| WORD | PROB. | WORD | PROB. | WORD | PROB. | WORD | PROB. |
| SPEECH | 0.1134 | PROBABILISTIC | 0.0778 | USER | 0.2541 | STARS | 0.0164 |
| RECOGNITION | 0.0349 | BAYESIAN | 0.0671 | INTERFACE | 0.1080 | OBSERVATIONS | 0.0150 |
| WORD | 0.0295 | PROBABILITY | 0.0532 | USERS | 0.0788 | SOLAR | 0.0150 |
| SPEAKER | 0.0227 | CARLO | 0.0309 | INTERFACES | 0.0433 | MAGNETIC | 0.0145 |
| ACOUSTIC | 0.0205 | MONTE | 0.0308 | GRAPHICAL | 0.0392 | RAY | 0.0144 |
| RATE | 0.0134 | DISTRIBUTION | 0.0257 | INTERACTIVE | 0.0354 | EMISSION | 0.0134 |
| SPOKEN | 0.0132 | INFERENCE | 0.0253 | INTERACTION | 0.0261 | GALAXIES | 0.0124 |
| SOUND | 0.0127 | PROBABILITIES | 0.0253 | VISUAL | 0.0203 | OBSERVED | 0.0108 |
| TRAINING | 0.0104 | CONDITIONAL | 0.0229 | DISPLAY | 0.0128 | SUBJECT | 0.0101 |
| MUSIC | 0.0102 | PRIOR | 0.0219 | MANIPULATION | 0.0099 | STAR | 0.0087 |
| AUTHOR | PROB. | AUTHOR | PROB. | AUTHOR | PROB. | AUTHOR | PROB. |
| Waibel_A | 0.0156 | Friedman_N | 0.0094 | Shneiderman_B | 0.0060 | Linsky_J | 0.0143 |
| Gauvain_J | 0.0133 | Heckerman_D | 0.0067 | Rauterberg_M | 0.0031 | Falcke_H | 0.0131 |
| Lamel_L | 0.0128 | Ghahramani_Z | 0.0062 | Lavana_H | 0.0024 | Mursula_K | 0.0089 |
| Woodland_P | 0.0124 | Koller_D | 0.0062 | Pentland_A | 0.0021 | Butler_R | 0.0083 |
| Ney_H | 0.0080 | Jordan_M | 0.0059 | Myers_B | 0.0021 | Bjorkman_K | 0.0078 |
| Hansen_J | 0.0078 | Neal_R | 0.0055 | Minas_M | 0.0021 | Knapp_G | 0.0067 |
| Renals_S | 0.0072 | Raftery_A | 0.0054 | Burnett_M | 0.0021 | Kundu_M | 0.0063 |
| Noth_E | 0.0071 | Lukasiewicz_T | 0.0053 | Winiwarter_W | 0.0020 | Christensen-J | 0.0059 |
| Boves_L | 0.0070 | Halpern_J | 0.0052 | Chang_S | 0.0019 | Cranmer_S | 0.0055 |
| Young_S | 0.0069 | Muller_P | 0.0048 | Korvemaker_B | 0.0019 | Nagar_N | 0.0050 |

Figure 3: An illustration of 4 topics from a 300-topic solution for the CiteSeer collection. Each topic is shown with the 10 words and authors that have the highest probability conditioned on that topic.

For the NIPS data set, 2000 iterations of the Gibbs sampler took 12 hours of wall-clock time on a standard PC workstation (22 seconds per iteration). Citeseer took 111 hours for 700 iterations (9.5 minutes per iteration). The full list of tables can be found at http://www.datalab.uci.edu/author-topic, for both the 100-topic NIPS model and the 300-topic CiteSeer model. In addition there is an online JAVA browser for interactively exploring authors, topics, and documents.

The results above use a single sample from the Gibbs sampler. Across different samples each sample can contain somewhat different topics i.e., somewhat different sets of most probable words and authors given the topic, since according to the author-topic model there is not a single set of conditional probabilities, $\theta$ and $\phi$, but rather a distribution over these conditional probabilities. In the experiments in the sections below, we average over multiple samples (restricted to 10 for computational convenience) in a Bayesian fashion for predictive purposes.

### 4.2 Evaluating predictive power

In addition to the qualitative evaluation of topic-author and topic-word results shown above, we also evaluated the proposed author-topic model in terms of perplexity, i.e., its ability to predict words on new unseen documents. We divided the $D = 1,740$ NIPS papers into a training set of $1,557$ papers with a total of $2,057,729$ words, and a test set of $183$ papers of which 102 are single-authored papers. We chose the test data documents such that each of the 2,037 authors of the NIPS collection authored at least one of the training documents.

Perplexity is a standard measure for estimating the performance of a probabilistic model. The perplexity of a set of test words, $(\mathbf{w}_d, \mathbf{a}_d)$ for $d \in \mathcal{D}^{\text{test}}$, is defined as the exponential of the negative normalized predictive likelihood under the model,

$$\text{perplexity}(\mathbf{w}_d|\mathbf{a}_d) = \exp\left[-\frac{\ln p(\mathbf{w}_d|\mathbf{a}_d)}{N_d}\right]. \quad (7)$$

Better generalization performance is indicated by a lower perplexity over a held-out document.

The derivation of the probability of a set of words given the authors is a straightforward calculation in the author-topic model:

$$\begin{aligned} p(\mathbf{w}_d|\mathbf{a}_d) &= \int d\theta \int d\phi\, p(\theta|\mathcal{D}^{\text{train}}) p(\phi|\mathcal{D}^{\text{train}}) \\ &\quad \times \prod_{m=1}^{N_d}\left[\frac{1}{A_d}\sum_{i\in \mathbf{a}_d, j}\theta_{ij}\phi_{w_m j}\right]. \end{aligned} \quad (8)$$

The term in the brackets is simply the probability for the word $w_m$ given the set of authors $\mathbf{a}_d$. We approximate the integrals over $\phi$ and $\theta$ using the point estimates obtained via Equations 5 and 6 for each sample of assignments $\mathbf{x}, \mathbf{z}$, and then average over samples. For documents with a single author this formula becomes

$$p(\mathbf{w}_d|a_d) = \frac{1}{S}\sum_{s=1}^{S}\prod_{m=1}^{N_d}\left[\sum_j \theta^s_{a_d j}\phi^s_{w_m j}\right],$$

where $\theta^s_{a_d j}$, $\phi^s_{w_m j}$ are point estimates from sample $s$, $S$ is the number of samples used, and $a_d$ is no longer a vector but a scalar that stands for the author of the document.

In the first set of experiments we compared the topic model (LDA) of Section 2.1, the author model of Section 2.2, and our proposed author-topic model from Section 2.3. For each test document, a randomly generated set of $N_d^{(train)}$ training words were selected and combined with the training data. Each model then made predictions on the other words in each test document, conditioned on the combination of both (a) the documents in the training data corpus and (b) the words that were randomly selected from the document. This simulates the process of observing some of the words in a document and making predictions about the rest. We would expect that as $N_d^{(train)}$ increases the predictive power of each model would improve as it adapts to the document. The author-topic and author models were both also conditioned on the identity of the true authors of the document. In all models, the topic and author distributions were all updated to new predictive distributions given the combination of the $N_d^{(train)}$ training words for the document being predicted and the full training data corpus. We averaged over 10 samples from the Gibbs sampler when making predictions for each word.

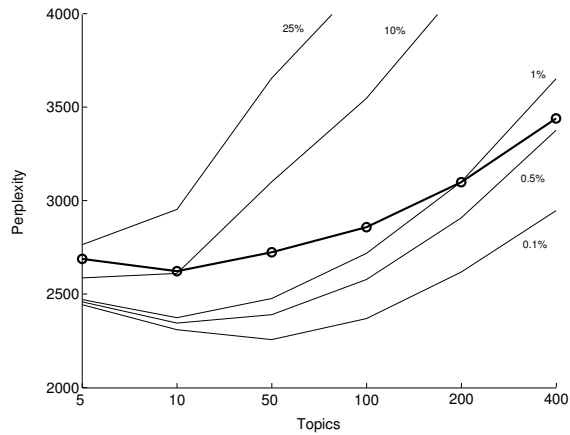

Figure 5: Perplexity of the 102 single-authored test documents from the NIPS collection, conditioned both on the correct author and authors ranked by perplexity using the model, as described in the text.

Figure 4 shows the results for the 3 models being compared. The author model is clearly poorer than either of the topic-based models, as illustrated by its high perplexity. Since a distribution over words has to be estimated for each author, fitting this model involves finding the values of a large number of parameters, limiting its generalization performance. The author-topic model has lower perplexity early on (for small values of $N_d^{(train)}$) since it uses knowledge of the author to provide a better prior for the content of the document. However, as $N_d^{(train)}$ increases we see a cross-over point where the more flexible topic model adapts better to the content of this particular document. Since no two scientific papers are exactly the same, the expectation that this document will match the previous output of its authors begins to limit the predictive power of the author-topic model. For larger numbers of topics, this crossover occurs for smaller values of $N_d^{(train)}$, since the topics pick out more specific areas of the subject domain.

To illustrate the utility of these models in predicting words conditioned on authors, we derived the perplexity for each of the 102 singled-authored test documents in the NIPS collection using the full text of each document and $S = 10$. The averaged perplexity as a function of the number of topics $T$ is presented in Fig-ure 5 (thick line). We also derived the perplexity of the test documents conditioned on each one of the authors from the NIPS collection, perplexity($\mathbf{w}_d|a$) for $a = 1, ..., K$. This results in $K = 2,037$ different perplexity values. Then we ranked the results and various percentiles from this ranking are presented in Figure 5. One can see that making use of the authorship information significantly improves the predictive log-likelihood: the model has accurate expectations about the content of documents by particular authors. As the number of topics increases the ranking of the correct author improves, where for 400 topics the averaged ranking of the correct author is within the 20 highest ranked authors (out of 2,037 possible authors). Consequently, the model provides a useful method for identifying possible authors for novel documents.

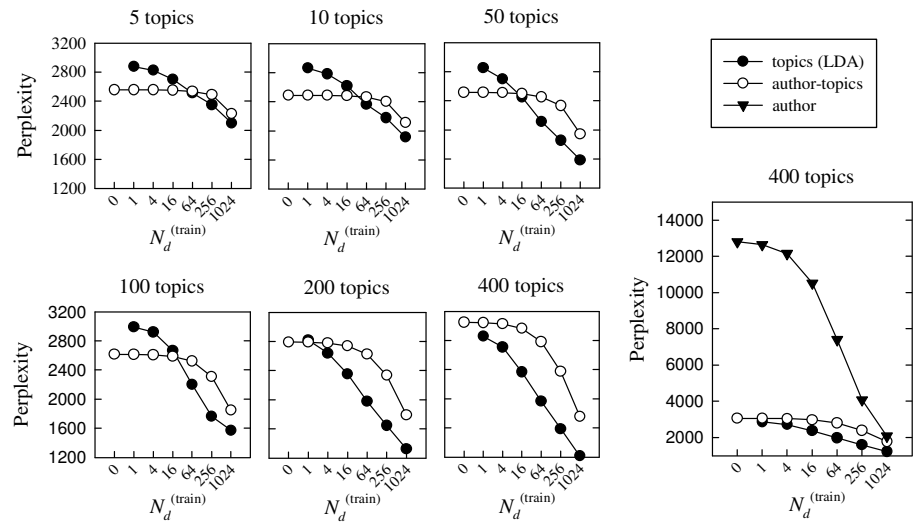

Figure 4: Perplexity versus $N_d^{(train)}$ for different numbers of topics, for the author, author-topic, and topic (LDA) models.

### 4.3 Illustrative applications of the model

The author-topic model could be used for a variety of applications such as automated reviewer recommendations, i.e., given an abstract of a paper and a list of the authors plus their known past collaborators, generate a list of other highly likely authors for this abstract who might serve as good reviewers. Such a task requires computing the similarity between authors. To illustrate how the model could be used in this respect, we defined the distance between authors $i$ and $j$ as the symmetric KL divergence between the topics distribution conditioned on each of the authors:

$$sKL(i,j) = \sum_{t=1}^{T} \left[ \theta_{it} \log \frac{\theta_{it}}{\theta_{jt}} + \theta_{jt} \log \frac{\theta_{jt}}{\theta_{it}} \right]. \quad (9)$$

As earlier, we derived the averaged symmetric KL divergence by averaging over samples from the posterior distribution, $p(\theta|\mathcal{D}^{\text{train}})$.

Table 1: Symmetric KL divergence for pairs of authors

| Authors | $n$ | T=400 | T=200 | T=100 |
|---|---|---|---|---|
| Bartlett_P (8) Shawe-Taylor_J (8) | - | 2.52 | 1.58 | 0.90 |
| Barto_A (11) Singh_S (17) | 2 | 3.34 | 2.18 | 1.25 |
| Amari_S (9) Yang_H (5) | 3 | 3.44 | 2.48 | 1.57 |
| Singh_S (17) Sutton_R (7) | 2 | 3.69 | 2.33 | 1.35 |
| Moore_A (11) Sutton_R (7) | - | 4.25 | 2.89 | 1.87 |
| MEDIAN | - | 5.52 | 4.01 | 3.33 |
| MAXIMUM | - | 16.61 | 14.91 | 13.32 |

Note: $n$ is number of common papers in NIPS dataset.

We searched for similar pairs of authors in the NIPS data set using the distance measure above. We searched only over authors who wrote more than 5 papers in the full NIPS data set—there are 125 such authors out of the full set of 2037. Table 1 shows the 5 pairs of authors with the highest averaged sKL for the 400-topic model, as well as the median and minimum. Results for the 200 and 100-topic models are also shown as are the number of papers in the data set for each author (in parentheses) and the number of co-authored papers in the data set (2nd column). All results were averaged over 10 samples from the Gibbs sampler.

Again the results are quite intuitive. For example, although authors Bartlett and Shawe-Taylor did not have any co-authored documents in the NIPS collec-

Table 2: Author entropies

| Author    | $n$ | T=400 | T=200 | T=100 |
|-----------|-----|-------|-------|-------|
| Jordan_M  | 24  | 4.35  | 4.04  | 3.61  |
| Fine_T    | 4   | 4.33  | 3.94  | 3.52  |
| Roweis_S  | 4   | 4.32  | 4.02  | 3.61  |
| Becker_S  | 4   | 4.30  | 4.06  | 3.69  |
| Brand_M   | 1   | 4.29  | 4.03  | 3.65  |
| MEDIAN    |     | 3.42  | 3.16  | 2.81  |
| MINIMUM   |     | 1.23  | 0.78  | 0.58  |

Note: $n$ is the number of papers by each author.

tion, they have in fact co-authored on other papers. Similarly, although A. Moore and R. Sutton have not co-authored any papers to our knowledge, they have both (separately) published extensively on the same topic of reinforcement learning. The distances between the authors ranked highly (in Table 1) are significantly lower than the median distances between pairs of authors.

The topic distributions for different authors can also be used to assess the extent to which authors tend to address a single topic in their work, or cover multiple topics. We calculated the entropy of each author's distribution over topics on the NIPS data, for different numbers of topics. Table 2 shows the 5 authors with the highest averaged entropy (for 400 topics) as well as the median and the minimum—also shown are the entropies for 200 and 100 topics. The top-ranked author, Michael Jordan, is well known for producing NIPS papers on a variety of topics. The papers associated with the other authors are also relatively diverse, e.g., for author Terrence Fine one of his papers is about forecasting demand for electric power while another concerns asymptotics of gradient-based learning. The number of papers produced by an author is not necessarily a good predictor of topic entropy. Sejnowski_T, for example, who generated the greatest number of papers in our NIPs collection, 37 of the training papers, is the 44th highest entropy author, with an entropy of 4.11 for $T = 400$.

## 5 Conclusions

The author-topic model proposed in this paper provides a relatively simple probabilistic model for exploring the relationships between authors, documents, topics, and words. This model provides significantly improved predictive power in terms of perplexity compared to a more impoverished author model, where the interests of authors are directly modeled with probability distributions over words. When compared to the LDA topic model, the author-topic model was shown to have more focused priors when relatively little is known about a new document, but the LDA model can better adapt its distribution over topics to the content of individual documents as more words are observed. The primary benefit of the author-topic model is that it allows us to explicitly include authors in document models, providing a general framework for answering queries and making predictions at the level of authors as well as the level of documents. Possible future directions for this work include using citation information to further couple documents in the model (c.f. Cohn & Hofmann, 2001), combining topic models with stylometry models for author identification, and applications such as automated reviewer list generation given sets of documents for review.

## Acknowledgements

The research in this paper was supported in part by the National Science Foundation under Grant IRI-9703120 via the Knowledge Discovery and Dissemination (KD-D) program. We would like to thank Steve Lawrence and C. Lee Giles for kindly providing us with the CiteSeer data used in this paper.